\documentclass[onecolumn,authoryear]{els-mrw} 

\usepackage{amsmath,amssymb,amsfonts,amsthm,makeidx,graphicx}
\usepackage{txfonts}
\usepackage{helvet}
\usepackage{aas_macros}

\begin{document}

\chapter{An observational overview of white dwarf stars}\label{chap1}

\author[]{Ingrid Pelisoli}
\author[]{Jamie Williams}%


\address[1]{\orgname{University of Warwick}, \orgdiv{Department of Physics}, \orgaddress{Gibbet Hill Road, Coventry, CV4 7AL, UK}}

\articletag{Chapter Article tagline: update of previous edition,, reprint..}

\maketitle

\begin{glossary}[Glossary]

\term{Astronomical unit:} distance scale corresponding to the average distance between the Earth and the Sun.

\term{Brown dwarf:} a substellar object whose mass is too low to ignite hydrogen fusion in its core, but high enough for deuterium fusion to occur.

\term{Gaia:} a space observatory managed by the European Space Agency (ESA) which is obtaining precise astrometric and photometric data for more than a billion sources.

\term{Main sequence star:} first stage of stellar evolution, where the star performs hydrogen fusion in its core.

\term{Giant star:} evolved star that has exhausted its core hydrogen and has increased in size.

\term{Gravitational waves:} perturbations of space-time due to the accelerated motion of bodies, such as binary stars.

\term{Parallax:} apparent angular displacement of an object when seen from two different positions.

\term{Photometry:} technique in which the light of an astronomical source is integrated regardless of wavelength (or within a wavelength range defined by a filter) providing a single measurement of flux.

\term{Proper motion:} apparent motion of an object in the plane of the sky compared to distant background sources.

\term{Spectroscopy:} technique in which the light of an astronomical source is split into its different wavelengths, providing the source's spectrum: flux as a function of wavelength.

\term{Spectral line:} flux absorption or emission at a specific wavelength corresponding to energy transitions of a given atom.

\end{glossary}

\begin{glossary}[Nomenclature]
\begin{tabular}{@{}lp{34pc}@{}}
$d$ & distance \\
$g$ & surface gravity \\
M$_{\odot}$ & solar mass unit: $1.98910\times10^{30}$~kg\\
$R$ & radius \\
pc & parsec, unit of length defined as the distance at which 1 astronomical unit subtends an angle of one arcsecond. \\
SDSS & Sloan Digital Sky Survey\\
SED & Spectral energy distribution\\
$T$ & temperature\\
$\theta$ & angular diameter \\
\end{tabular}
\end{glossary}

\begin{abstract}[Abstract]

White dwarf stars are the most common final stage of stellar evolution. Since the serendipitous discovery of the first white dwarf by William Herschel and the first physical models by Subrahmanyan Chandrasekhar and Arthur Eddington, there have been a lot of advances in the field fueled by new observational data. With new astrometric measurements enabling us to identify hundreds of thousands of white dwarf candidates, and spectroscopic surveys revealing a plethora of chemical elements in white dwarf atmospheres pointing at spectral evolution and interaction with planetary bodies, there is a lot we can learn from the characterization of observed white dwarfs.

This chapter provides an observational overview of white dwarf stars, describing how they are identified and characterized, and the main properties of the observed population. 

\textbf{Keywords}: Stellar evolution: Late stellar evolution -- Stellar remnants: Compact objects: White dwarf stars
\end{abstract}

\begin{glossary}[Key Points]
\begin{itemize}
    \item White dwarf stars are the most common outcome of stellar evolution.
    \item Spectroscopy is a powerful tool for characterizing white dwarf stars.
    \item Asteroseismology is a technique used to infer the interior structure of stars, including white dwarfs.
    \item Planetary material, mainly in the form of debris, has been detected around a large fraction of white dwarfs.
\end{itemize}
\end{glossary}

\section{Introduction}\label{sec:intro}

The mass distribution of a newborn stellar population can be described by an initial-mass function. There are many possible approximations for this function in the literature \citep[e.g.][]{Salpeter1955}, all of which agree in one important aspect: the vast majority of stars ($>90$\%) are born with masses below 10~M$_{\odot}$. The beginning of the life of any star is marked by the nuclear fusion of hydrogen into helium --- by definition, all stars are capable of performing this nuclear reaction at the start of their lives. Following a brief transition phase during which the inert helium core contracts and the outer layers of the star expand, the next stage in the life of a star is typically marked by helium fusion, although only stars with initial masses above about 0.45~M$_{\odot}$ can reach temperature and pressure conditions in their core to burn helium into carbon and oxygen. This is where nuclear fusion stops for stars with masses below 10~M$_{\odot}$, with only the most massive ones within this range also consuming carbon and forming cores composed of oxygen, neon, and magnesium. Once nuclear fusion halts, the associated release of energy, which counteracted gravity and prevented the stellar core from collapsing under its own weight, ceases. The core thus contracts until a new source of internal pressure prevents further collapse: electron-degeneracy pressure. As dictated in the Fermi Exclusion Principle, two electrons cannot occupy the exact same energy state, and as a result will exercise pressure when confined enough. This electron-degenerate core resulting from stellar evolution is surrounded by an envelope composed of any hydrogen and helium not consumed during the evolution. This leftover fossil of stellar evolution has a name: a white dwarf star \citep[see][for an extensive review]{Koester2013}.

White dwarf stars are thus the most common final stage of stellar evolution. Their compact cores have sizes comparable to the Earth, but extremely strong gravities that can be up to almost 10 million times stronger than the Earth's gravity. As a result, white dwarfs have stratified structures as heavier materials sink to the core whereas lighter ones float to the atmosphere. A typical white dwarf has a carbon-oxygen core, surrounded by a helium mantle and, finally, by a hydrogen layer. Over 99\% of a white dwarf's mass is in its core, but the light helium and hydrogen layers contribute significantly to the star's radius. Lower mass white dwarfs can have helium cores, or hybrid helium/carbon cores, whereas the most massive systems can have oxygen/neon/magnesium cores. The atmosphere's composition can also vary depending on evolution and on interaction with companion stars or planetary bodies.

White dwarf stars play an important role in many aspects of astronomy and astrophysics. As a last stage in the life of a star, their properties reflect and can be used to constrain the many different physical process that happen during stellar evolution. As old objects, they can be used to date stellar populations in a field known as cosmochronology. Their pristine atmospheres can reveal the internal compositions of planetary bodies that are accreted by the white dwarf. Their extreme properties, in particular high gravities, allow for tests of fundamental physics that would be difficult to carry out on Earth. All of these possible applications of white dwarf stars depend on an accurate identification and characterization of white dwarfs from observational data.

\section{Identifying white dwarf stars in observational data}


Although technically the first white dwarf was found by William Herschel in 1785, its nature was not clarified until almost 150 years later. 
The white dwarf observed by Herschel was 40 Eridani B, which is part of the triple system 40 Eridani \citep{Herschel1785}. Its nature came under attention with the advent of spectroscopy towards the end of the of the 19th and beginning of 20th centuries, when it was discovered that 40 Eridani B was hydrogen-dominated (i.e. a white in appearance, A-type star). Up until then, all A-type stars observed were main sequence stars, which are still in the hydrogen-burning phase and are much brighter compared to the dim 40 Eridani B. Willem Luyten is believed to have coined the term "white dwarf" to refer to these faint A-type stars \citep{Luyten1922}. The dimness of white dwarf stars, which required them to have a much smaller radius than a typical A-type star and thus seemingly unachievable densities, remained a puzzle until 1926, when R. H. Fowler realized that when quantum mechanics, then at its infancy, was taken into account, the high densities were not a problem as electron-degeneracy pressure was enough to maintain hydrostatic equilibrium \citep{Fowler1926}. This idea was further developed by Subrahmanyan \citet{Chandrasekhar1935} and Arthur \citet{Eddington1935}, in particular with the inclusion of relativistic effects.

Interest in white dwarfs continued after the first identifications, and by 1950 over a hundred white dwarfs were known, mainly due to efforts led by Willem Luyten \citep{Luyten1950}. As 40 Eridani B, these were all identified because they showed spectral types typical of bright stars, but were themselves faint despite their presumably close distances given their detectability. This same method is still used until today with modern instruments and more precise data. 

The first comprehensive catalog of white dwarf stars was published by \citet{Eggen1965}, containing the characterization of 166 white dwarfs. This was followed by seminal works by George McCook and Edward Sion, who published a series of catalogs starting in 1977, the last of which was published in 1999 and contained 2249 white dwarfs \citep{McCookSion1999}. A large fraction of those were discovered as part of the Palomar-Green survey \citep{PGsurvey}, which obtained spectroscopic observations of objects showing strong blue or ultraviolet colors. Despite the name ``white" dwarf, which as explained above was motivated by the initial discoveries, bright white dwarf stars more typically have blue colors.

The beginning of the 21st century brought a revolution to the field with the start of the Sloan Digital Sky Survey \citep[SDSS,][]{York2000}. The main focus of SDSS then was extragalactic, as they attempted to understand the three-dimensional structure of the Universe by identifying distant sources. Part of their goal was to find quasars, the bright blue cores of distant galaxies. This was done by first imaging the sky and then obtain spectroscopic follow-up observations for objects obeying certain color selection criteria. Luckily, white dwarf stars have similar colors to quasars, and as a result tens of thousands of them were observed serendipitously. By the data release 16, the last one in SDSS's phase IV before its aims were significantly expanded in scope (including the creation of a dedicated white dwarf survey), over 35,000 white dwarf stars had been identified through SDSS spectra \citep{Kepler2021}.

The next major revolution in the field came with the {\it Gaia} satellite \citep{Gaia}, which provided for the first time astrometric measurements on a large scale for white dwarfs. Before {\it Gaia} data became available, the identification of white dwarf stars relied on finding intrinsically faint galactic blue sources. However, determining whether an apparently faint object is intrinsically faint is not straightforward. Ideally, the distance to the source must be known, but distances are remarkably difficult to determine in astronomy. The most accurate method is astrometry, whereby the parallax is measured and, through trigonometry, provides a distance estimate. In 2017, only a very small fraction of white dwarfs, less than 250 out of tens of thousands known, had parallax measurements available \citep{Bedard2017}. Instead, proper-motion criteria were typically employed. On average, distant stars will show smaller proper motion than nearby stars and, based on this principle, high proper-motion was used as a proxy for nearby distances. This method can lead to contamination from stars that move fast for other reasons, such as ejection from a binary system.

\begin{figure}[h]
\centering
\includegraphics[width=0.7\textwidth]{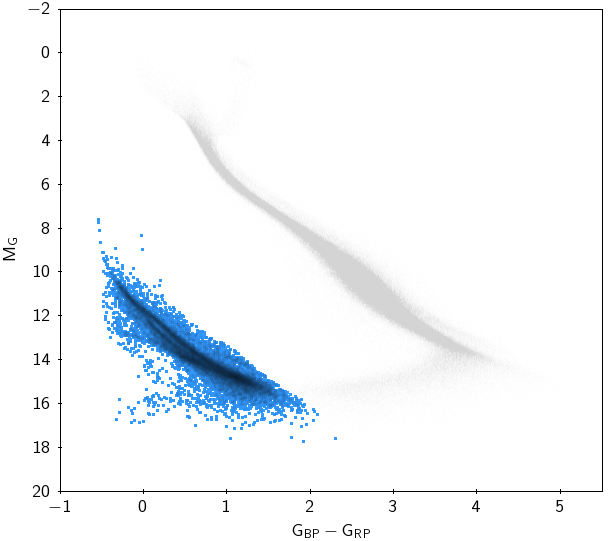}
\caption{The {\it Gaia} absolute $G$ magnitude ($M_G$) --- a proxy for intrinsic brightness --- as a function of the $G_{BP} - G_{RP}$ color --- a proxy for temperature. The high-probability white dwarf candidates identified in {\it Gaia} data release 3 are shown in blue. The gray dots show the {\it Gaia} Catalog of Nearby Stars \citep{GCNS}. Thanks to the precise parallaxes provided by {\it Gaia}, the intrinsic dimness of white dwarfs can be easily identified and their cooling sequence is revealed to unprecedented detail.}
\label{fig:gaiahr}
\end{figure}

The first data release of {\it Gaia} did not improve considerably on the situation, providing only six direct parallax measurements of white dwarfs and additional 46 indirect measurements, where the parallax of a brighter star gravitationally bound to the white dwarf in a binary system could be used \citep{Tremblay2017}. With a longer baseline, data release 2 was truly groundbreaking, providing parallax estimates for 1.3 billion sources. This allowed for the first time the identification of white dwarf stars by directly detecting their low intrinsic brightness on a large scale. In data release 2, 260,000 high-probability white dwarf candidates were identified, potentially increasing the number of known white dwarf stars by more than seven-fold \citep{Fusillo2019}. In data release 3, the number of high-probability white dwarf candidates increased to 359,000 \citep{Fusillo2021}. These identifications are shown in Figure~\ref{fig:gaiahr}.

\section{Spectroscopic characterization of white dwarf stars}

Any white dwarf identification method, including astrometry, typically requires spectroscopic follow-up to confirm the candidates, as inaccurate measurements can lead to other objects, such as quasars, being misclassified as white dwarfs, sometimes even with very high white dwarf probability estimates. Spectroscopy also provides the opportunity for further characterization, including the determination of the atmosphere composition, estimation of effective temperature and surface gravity, and detection of magnetic fields.

\subsection{Spectroscopic classes}

\begin{figure}[h]
\centering
\includegraphics[width=0.8\textwidth]{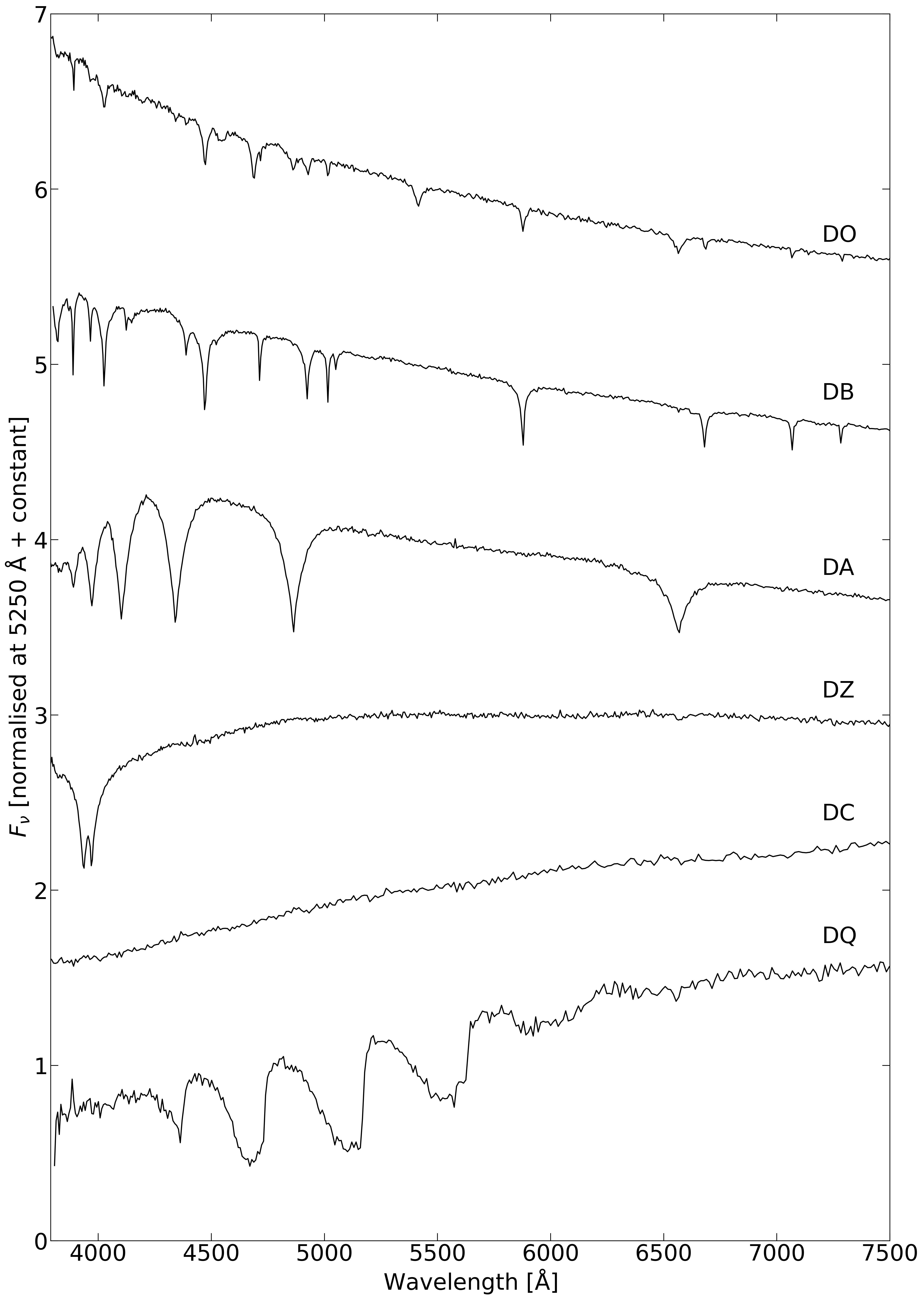}
\caption{White dwarf spectra from SDSS representing the main white dwarf spectroscopic classes.}
\label{fig:spec}
\end{figure}

White dwarfs are placed into different spectroscopic classes according to the elements that are visible in their spectra. Although somewhat subjective and strongly dependent on the quality of the data, this method is still largely used to date. The spectral classes of white dwarf stars are preceded by a ``D'' to represent the degenerate nature \citep[as originally proposed by][]{Luyten1945}, followed by one or more letters describing which elements are visible in the spectrum. The main white dwarf classes --- A, B, and O --- are analogous to the main sequence OBA types, where A-type show hydrogen spectral lines, B-type show neutral helium lines, and O type show ionized helium lines. Other spectral types have been defined for white dwarfs: C refers to white dwarfs with continuum spectra and no visible lines, Q to white dwarfs showing carbon lines or bands, and Z to white dwarfs showing other metal enhancement\footnote{In astronomy, all elements other than hydrogen and helium are referred to as metals.}. These main spectral classes are illustrated in Figure~\ref{fig:spec}. Additionally, when magnetic fields are visible in the spectra through Zeeman splitting of spectral lines, an ``H'' is added to the end of the class. If emission lines are present, an ``e'' is added. A summary of white dwarf spectral classes and their characteristics is given in Table~\ref{chap1:tab1}. In general, different combinations of letters can be used to describe all visible spectral characteristics, with the most prominent chemical element written first. For example, a DAZ white dwarf would show strong hydrogen lines and traces of metals, whereas a DZA would show strong metal lines and weak hydrogen lines. This classification scheme was developed by \citet{Sion1983}.

\begin{table}[t]
\TBL{\caption{White dwarf spectroscopic classes and their characteristics.}\label{chap1:tab1}}
{\begin{tabular*}{\textwidth}{@{\extracolsep{\fill}}@{}lll@{}}
\toprule
\multicolumn{1}{@{}l}{\TCH{Class}} &
\multicolumn{1}{l}{\TCH{Characteristic}} &
\multicolumn{1}{l}{\TCH{Typically strongest optical line}}\\
\colrule
DO & ionised He lines & 4686~$\AA$ \\
DB & neutral He lines & 4471$~\AA$ \\
DA & H lines & 6563$~\AA$ (H$\alpha$) \\
DC & continuum, no lines & -- \\
DQ & C lines or C$_2$ bands & 4367~$\AA$ or Swan bands \\
DZ & metal lines & 3968 and 3934~$\AA$ (Ca H and K) \\
\botrule
\end{tabular*}}

\end{table}

The majority of white dwarf stars ($>60$\%) are of DA spectral type. That is unsurprising: as previously mentioned, the strong gravity of white dwarfs causes heavier elements to sink, leaving the lightest elements in the outer layers, which are directly observed through spectroscopy. Hence hydrogen, being the lightest element, is the one to be more commonly observed in the outer layers. The DC spectral type is also easy to understand: below $\approx 11,000$~K the energy in the atmosphere is too low to excite any transitions in the helium atom, and the same happens below 5,000~K for hydrogen. Therefore, DB white dwarfs turn into DCs below $\approx 11,000$~K, and DA white dwarfs turn into DCs below $\approx 5,000$~K. The DO and DB types are also connected: helium gets ionized above $\approx 45,000$~K, hence a helium-dominated white dwarf appears as a DO above $\approx 45,000$~K, and as a DB below that when helium recombines.

The existence of helium dominated atmospheres is in turn explained by a couple of different processes. First, in some cases the hydrogen can be almost completely consumed during evolution as a result of a late event of helium burning (called a late helium flash),
which triggers an extensive convection zone that mixes any surface hydrogen into the core, where it gets almost entirely burned. These white dwarfs are born with helium-, carbon-, and oxygen-rich atmospheres and are sometimes referred to as PG 1159 stars, after their prototype system. With time, the carbon and oxygen sink down due to gravitational settling, leaving a helium-rich atmosphere that appears as a DO or DB depending on temperature. Second, at the beginning of the white dwarf cooling sequence the outer layers are not yet stratified, and helium and hydrogen are mixed. Helium is typically a hundred times more abundant (in mass) than hydrogen at the beginning of the white dwarf cooling sequence and hence the hydrogen is essentially invisible in this mixed atmosphere. With time, gravitational settling triggers stratification of the atmosphere, and the hydrogen lines become visible. These stars then start as DO type and evolve to DA type. It is worth noting that the hydrogen might be mixed again as the white dwarf cools down and a convection layer forms, leading these hydrogen dominated atmospheres to become helium-dominated again \citep{Bedard2022}.

The presence of carbon in the atmosphere can be explained by two different processes. First, as white dwarfs cool down, energy transport by radiation alone becomes insufficient to maintain thermal equilibrium, and a convection layer that transports both energy and material forms. For helium-dominated atmospheres the convective layer reaches the degenerate core of the white dwarf around $12,000$~K and brings carbon to the atmosphere. When there is sufficient carbon, the white dwarf will turn into a DQ (typically this happens when the white dwarf is already cool enough that no hydrogen or helium lines are visible). However, this does not explain the existence of DQ white dwarfs with hotter temperatures, sometimes called hot or warm DQs, which typically have temperatures above $\approx 10,000$~K. These hot DQs are attributed to stellar mergers, where two stars merge into one and the core contents, which are carbon rich, are mixed into the atmosphere.

Finally, the existence of DZ white dwarfs is explained by an extrinsic rather than intrinsic process. Like our Sun, other stars are orbited by planets and, likely, also asteroids and comets. These bodies, either due to stellar evolution or due to dynamical interaction with other nearby bodies, can be accreted onto the atmosphere of the white dwarf leading to metal enhancement (also called metal pollution). The sinking times of metals are, at most, a few million years, which is very quick compared to the cooling ages of white dwarfs which are on the order of billions of years. This implies that if metals were accreted in a single event, their observation would be extremely unlikely given their fast sinking times. As that is not the case --- metal enhancement is often observed ($> 10$\% of white dwarfs) --- we can infer that DZ white dwarf stars experience continuous accretion.

\subsection{Temperatures and masses}

The physical properties of white dwarf stars can be derived by two main methods: fitting of the spectral energy distribution (SED) and fitting of spectra. Both methods rely on spectral models derived from atmosphere models, which attempt to describe the energy transport in the atmosphere and resulting emission profile based on a number of physical inputs, including element abundances and opacities \citep{Koester2010}. In SED fitting, photometric measurements in different filters are compared to predicted synthetic photometry derived from models. The typical output of an SED fit is the white dwarf temperature and the angular diameter, $\theta \propto R^2/d^2$, which allows the radius to be obtained if the distance is known. To first order, the temperature is what determines the slope of the SED, whereas the angular diameter determines the normalization. In the spectral fitting method, the observed spectrum (continuum and visible lines) is fit with spectral models. The typical output is the temperature and surface gravity (expressed as $\log~g$, where $g$ is the gravity measured in cgs units: cm/s). The temperature is primarily determined by the depth of the spectral lines, whereas the gravity affects their width, though the effects are correlated.

Thanks to their degenerate nature, where pressure is a function of density rather than temperature, the masses and radii of white dwarf stars are related. This mass-radius relationship depends on core composition and shows scatter due to varying atmosphere thickness, but it is roughly represented by $R \propto M^{-1/3}$. This relationship means, in short, that the higher the white dwarf mass, the smaller its radius. Importantly, it means that both the radii obtained via the SED method and the surface gravities obtained via the spectroscopic method can lead to mass estimates.

\begin{figure}[h]
\centering
\includegraphics[width=0.6\textwidth]{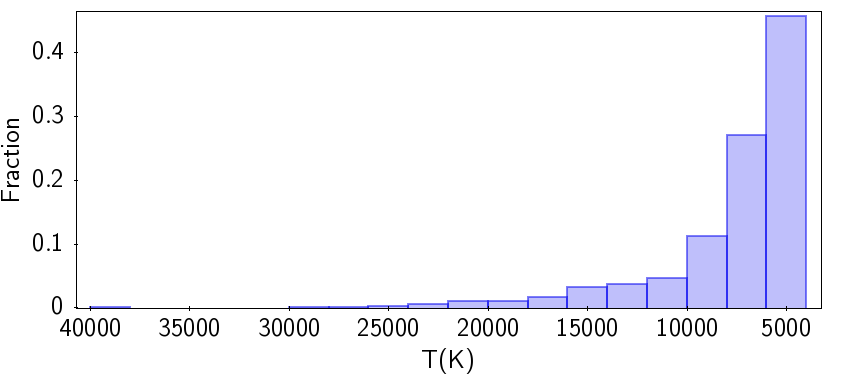}
\caption{The fraction of white dwarfs in each temperature bin for white dwarfs within 40~pc of the Sun. Note that temperature decreases to the right, as is the convention is astronomy. It is evident that hot white dwarfs are very rare, whereas cool white dwarfs are abundant given that white dwarfs spend longer times at cooler temperatures.}
\label{fig:Thist}
\end{figure}

White dwarf stars are observed with temperatures ranging from hundreds of thousands to only a few thousand Kelvin. The initial cooling is relatively fast and progressively slows down as the luminosity of the white dwarf decreases. The cooling rate increases again when the white dwarf has cooled enough that its core turns into a crystal (in essence, it freezes). It takes only one billion years for a white dwarf's temperature to decrease from 100,000~K to 10,000~K, but five times that for it to decrease from 10,000~K to 5,000~K. As a consequence, hot white dwarf stars are less common than cooler white dwarfs, simply because the timescale for them to cool down is fast. This is illustrated in Figure~\ref{fig:Thist}.

\begin{figure}[h]
\centering
\includegraphics[width=0.6\textwidth]{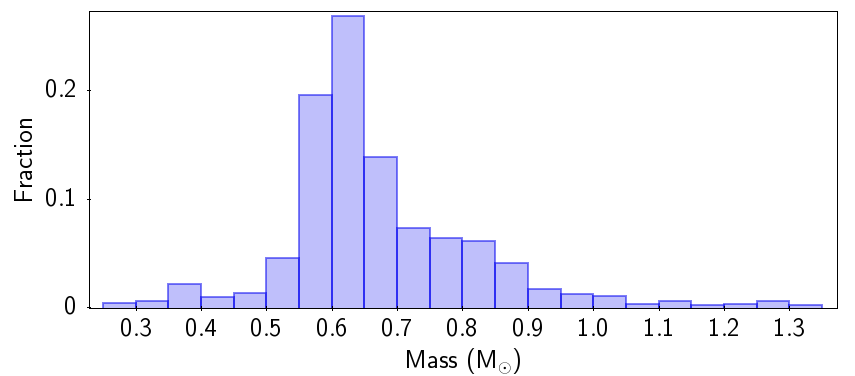}
\caption{The fraction of white dwarfs in each mass bin for white dwarfs within 40~pc of the Sun. There is a dominant peak at $0.6$~M$_{\odot}$, with a secondary peak around $0.4$~M$_{\odot}$ and an excess around $0.8$~M$_{\odot}$.}
\label{fig:Mhist}
\end{figure}

The white dwarf mass distribution is in turn more complex and not monotonic, as shown in Figure~\ref{fig:Mhist}. The average (and mode) mass of white dwarf stars, also known as canonical mass, is around $0.6$~M$_{\odot}$. This peak can simply be explained by normal stellar evolution. The time it takes for a star to exhaust its core fuel and turn into a white dwarf scales with its mass: the higher the mass, the faster the fuel is consumed. Given the age of our Galaxy, stars with masses above around 1.0~M$_{\odot}$ have had time to evolve off the main sequence. A considerable amount of mass is lost during stellar evolution, such that these main sequence stars form white dwarfs with masses around 0.6~M$_{\odot}$\footnote{The relationship between a star's initial mass and the mass of the resulting white dwarf is known as initial-to-final mass relationship.}. Part of the high-mass tail in this histogram can also be explained by stellar evolution, as higher mass stars that turn into higher mass white dwarfs have also had time to evolve. A fraction of higher mass white dwarfs are explained by stellar mergers -- exactly how much is still undetermined due to uncertainties on the initial-mass function of stars. The low-mass end of the white dwarf mass distribution also requires binary evolution to be explained. Only stars with masses above 0.45~M$_{\odot}$ have had time to evolve off the main sequence within the age of the Universe --- any white dwarf with a mass below that requires some form of enhanced mass loss to exist. This enhanced mass loss is explained, in most if not all cases, by binary interaction. It is worth mentioning that about half of stars are in binary systems (though this fraction is strongly dependent on stellar mass), hence these contributions of binary stellar evolution to the white dwarf population are not surprising. No white dwarfs are observed above 1.4~M$_{\odot}$, which is in agreement with theoretical expectations: this is the so-called Chandrasekhar limit, above which electron-degeneracy pressure would not be enough to sustain hydrostatic equilibrium, and the star would collapse to form a neutron star or black hole.

\subsection{Magnetic fields}

\begin{figure}[h]
\centering
\includegraphics[width=0.7\textwidth]{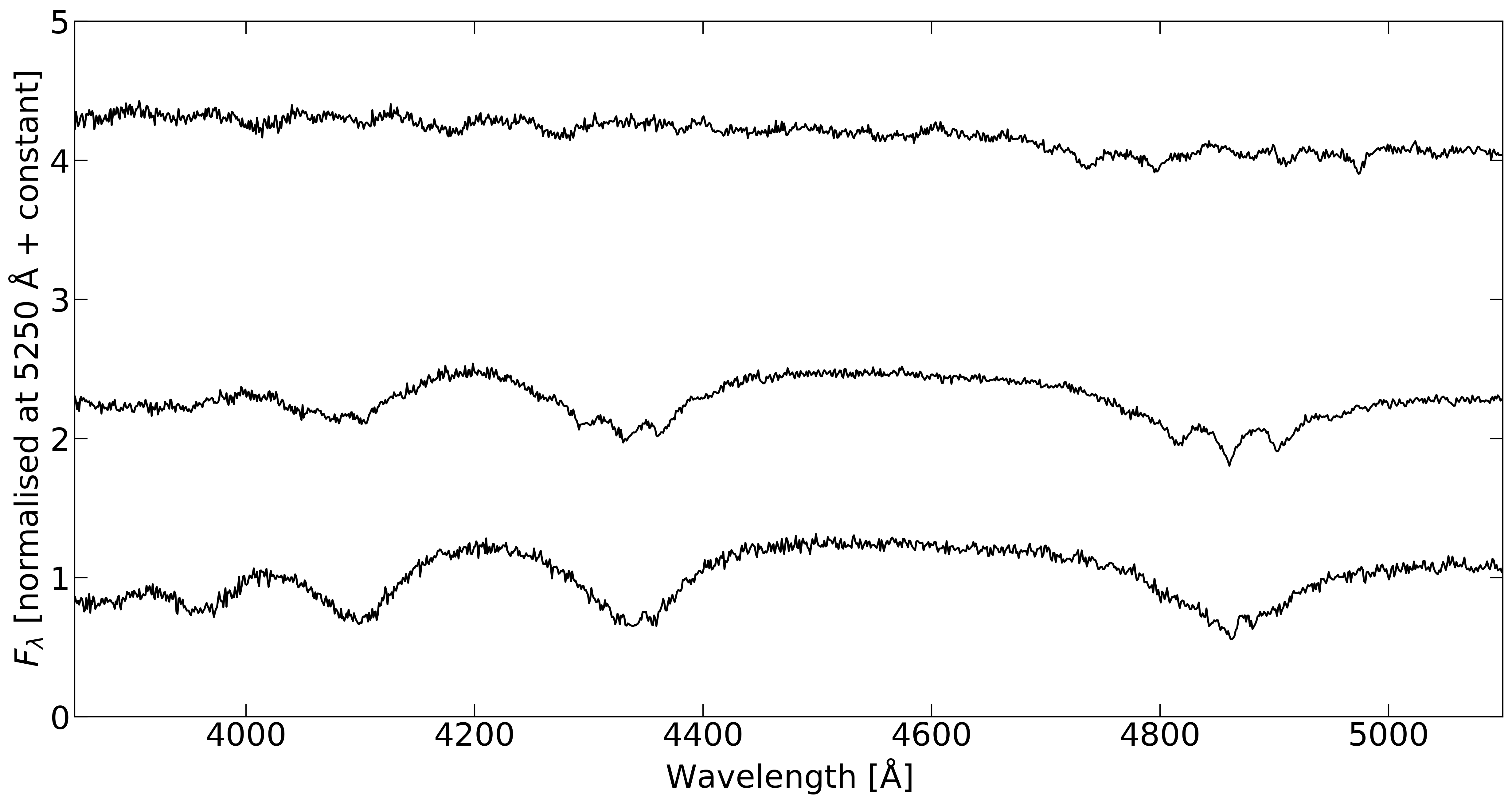}
\caption{Spectra of three magnetic white dwarf stars of spectral type DAH. The estimated magnetic fields are 2, 7 and 43~MG from bottom to top.}
\label{fig:Bspecs}
\end{figure}

Another characteristic of white dwarf stars that can sometimes be revealed by spectroscopy is the presence of magnetic fields \citep[see the review by][]{Ferrario2015}. A magnetic field changes the energy levels in an atom by lifting a degeneracy (on the conveniently
named quantum magnetic number) such that transitions that before had the same energy, and as a consequence contributed to the same spectral line, show different energies, leading to the spectral line being split into components. The level of splitting and even the number of components depends on the intensity of the magnetic field. In white dwarf stars, magnetic fields from about 40~kG (4~T) to 700~MG (70000~T) have been detected for about 20\% of white dwarfs. For the typical spectroscopic resolutions of surveys such as SDSS, fields with strengths above 1~MG lead to detectable line splitting (see examples in Figure~\ref{fig:Bspecs}). Fields with strengths lower than that require other techniques to be detected, in particular polarimetry, where the polarization of light caused by a magnetic field is used to infer its presence.

The origin of these sometimes very strong magnetic fields in white dwarfs (as a comparison, the Sun's magnetic field is around 1~G or 0.0001~T) is not fully understood. There are three main scenarios proposed to explain their existence, and likely they all contribute to the observed population to different degrees. First, some magnetic fields might simply be so-called fossil fields, where the field has existed before the white dwarf phase and is amplified when the core contracts. Magnetic flux is proportional to the inverse of the stellar radius squared ($\propto 1/R^2$), such that a factor of a 100 decrease in radius (which is what will happen when the Sun turns into a white dwarf) will result in a 10,000 factor increase in the magnetic field. This could potentially explain fields up to a few MG, but stronger fields are harder to explain with this scenario. The high fraction of white dwarfs that show magnetic fields is also not consistent with a fossil-field scenario only. Another possible mechanism is stellar mergers. During mergers, the movement of charged particles leads to a dynamo effect that can generate stronger magnetic fields. This scenario is favored by the fact that magnetic white dwarf stars do show an average mass higher than $0.6$~M$_{\odot}$, which could be explained by a fraction of them coming from mergers that form more massive white dwarfs. Finally, a third scenario is that the magnetic field forms or at least emerges to the surface only during the white dwarf phase. One motivation behind this hypothesis is that the fraction of magnetic white dwarf stars increases with age, suggesting a mechanism that is age dependent or, since the evolution of white dwarfs is a cooling process, more efficient at cooler temperatures. One possibility is a core-convection-driven dynamo similar to the one in operation in the Earth, as crystallization in the core as white dwarfs cool down leads to element separation which emulates convection, but whether such a dynamo would be efficient enough to produce magnetic fields with the observed strength is a subject of ongoing research.

\section{Asteroseismology and the interiors of white dwarf stars}

Whereas spectroscopy allows one to probe into the atmospheres of white dwarfs, which are directly observable, it does not allow for any direct measurement of the interiors of white dwarf stars. Luckily, there is another technique the enables one to probe into the internal structure of white dwarfs: asteroseismology, or the study of stellar pulsations \citep[see][for a thorough review of white dwarf pulsations]{Corsico2019}.

As white dwarfs cool down, the ionization ratios in their interiors change. The most remarkable change happens at around $12,000$~K, when hydrogen becomes partially ionized. This changes the opacity of internal layers and affects energy transport, resulting in temperature oscillations on the white dwarf surface, which in turn cause flux variability, or pulsations. Because these pulsations are a direct consequence of changes in the internal layers of the white dwarf, modeling the pulsations allows the internal structure of the white dwarf to be modeled. The observed pulsations are compared to characteristic pulsation modes calculated for different models of white dwarf interiors, and the best match reveals the most likely internal structure of the star.

\begin{figure}[h]
\centering
\includegraphics[width=0.7\textwidth]{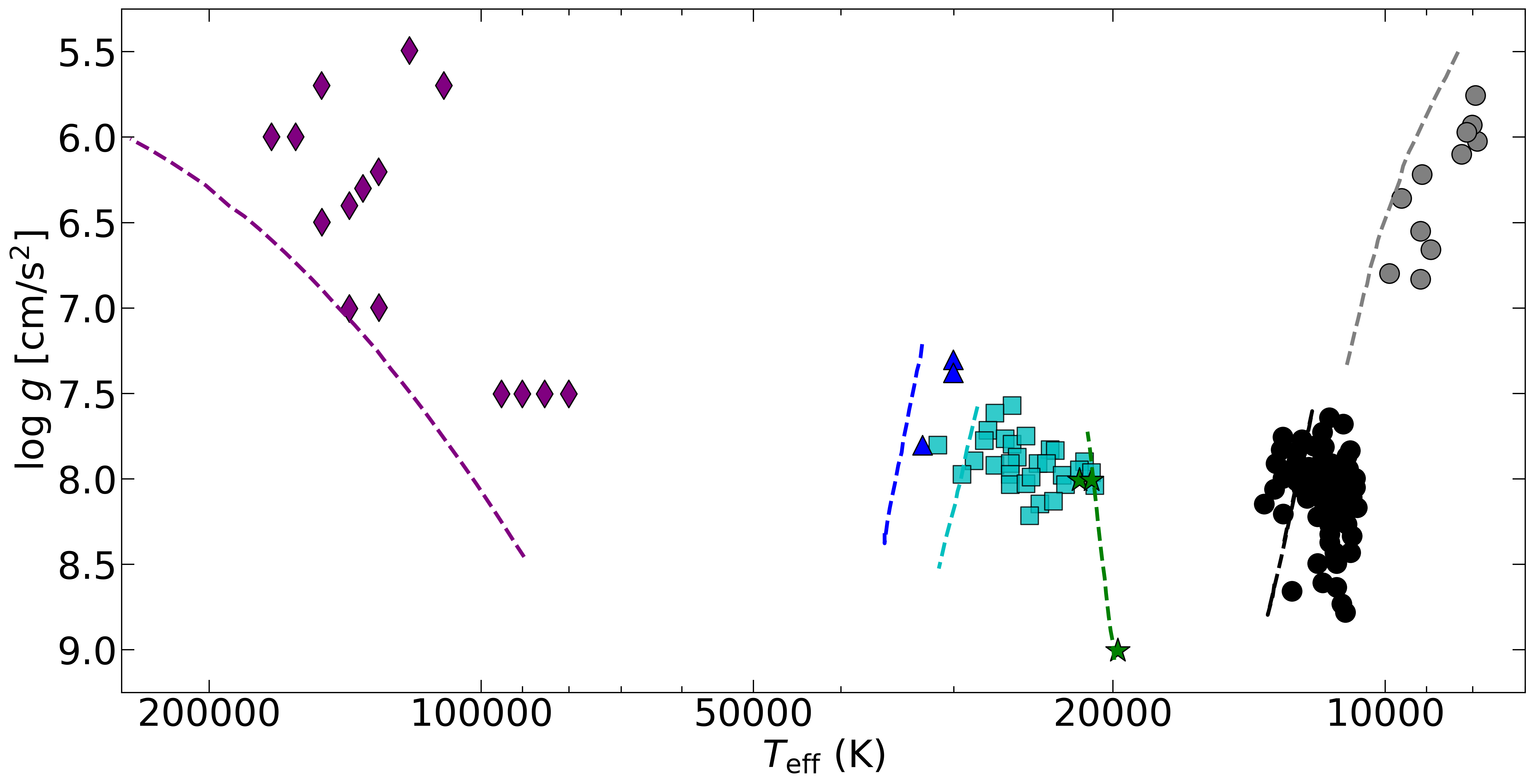}
\caption{Main instability strips for white dwarfs. The dashed lines indicated the theoretical upper edge of each strip; pulsations are predicted to occur for temperatures lower than this edge. Purple diamonds are DOVs, blue triangles are hot DAVs, cyan squares are DBVs, green stars are DQVs, black and gray circles are DAVs (low-mass systems with longer periods are shown in gray).}
\label{fig:strips}
\end{figure}

\begin{figure}[h]
\centering
\includegraphics[width=0.66\textwidth]{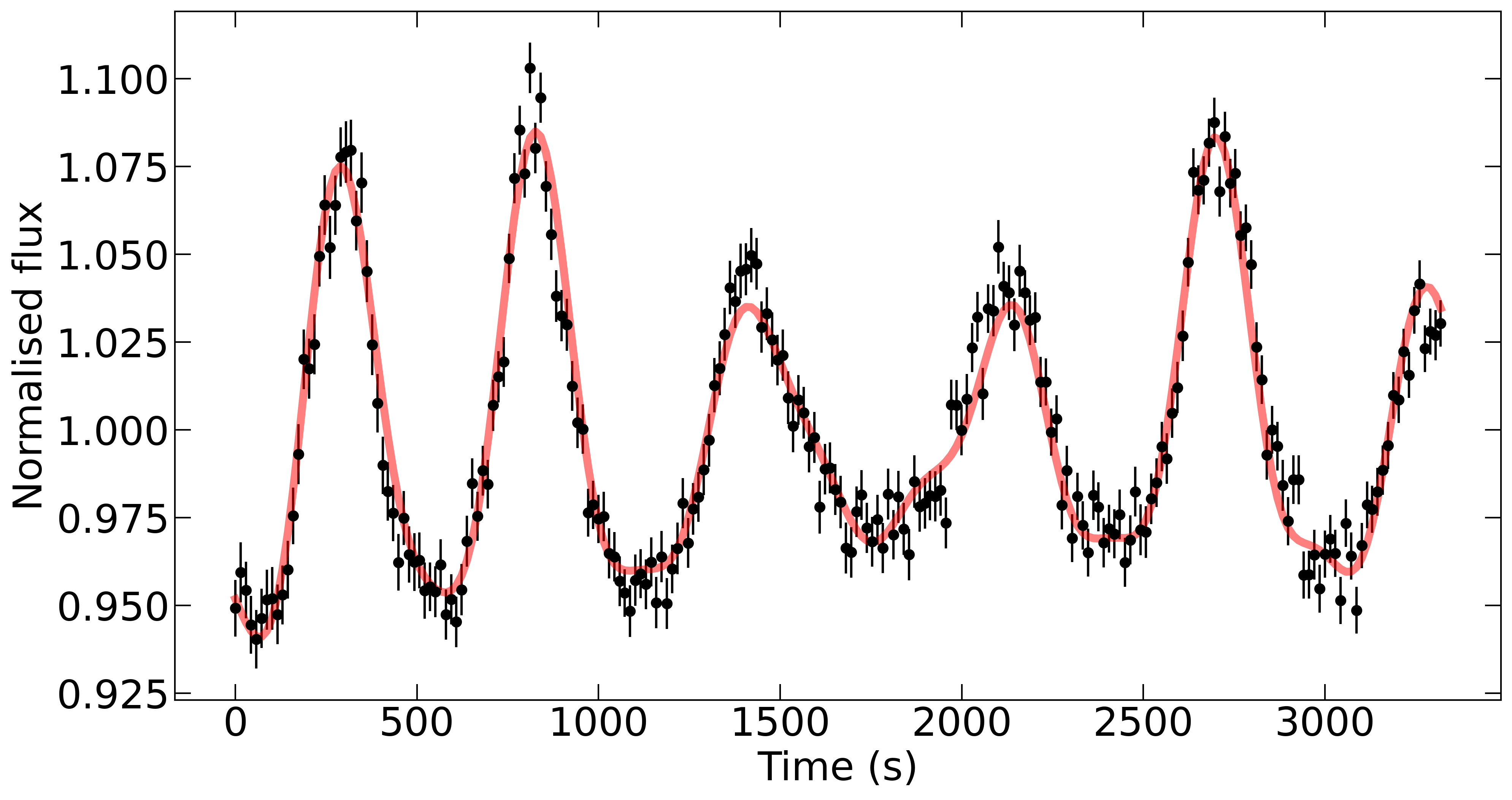}
\includegraphics[width=0.33\textwidth]{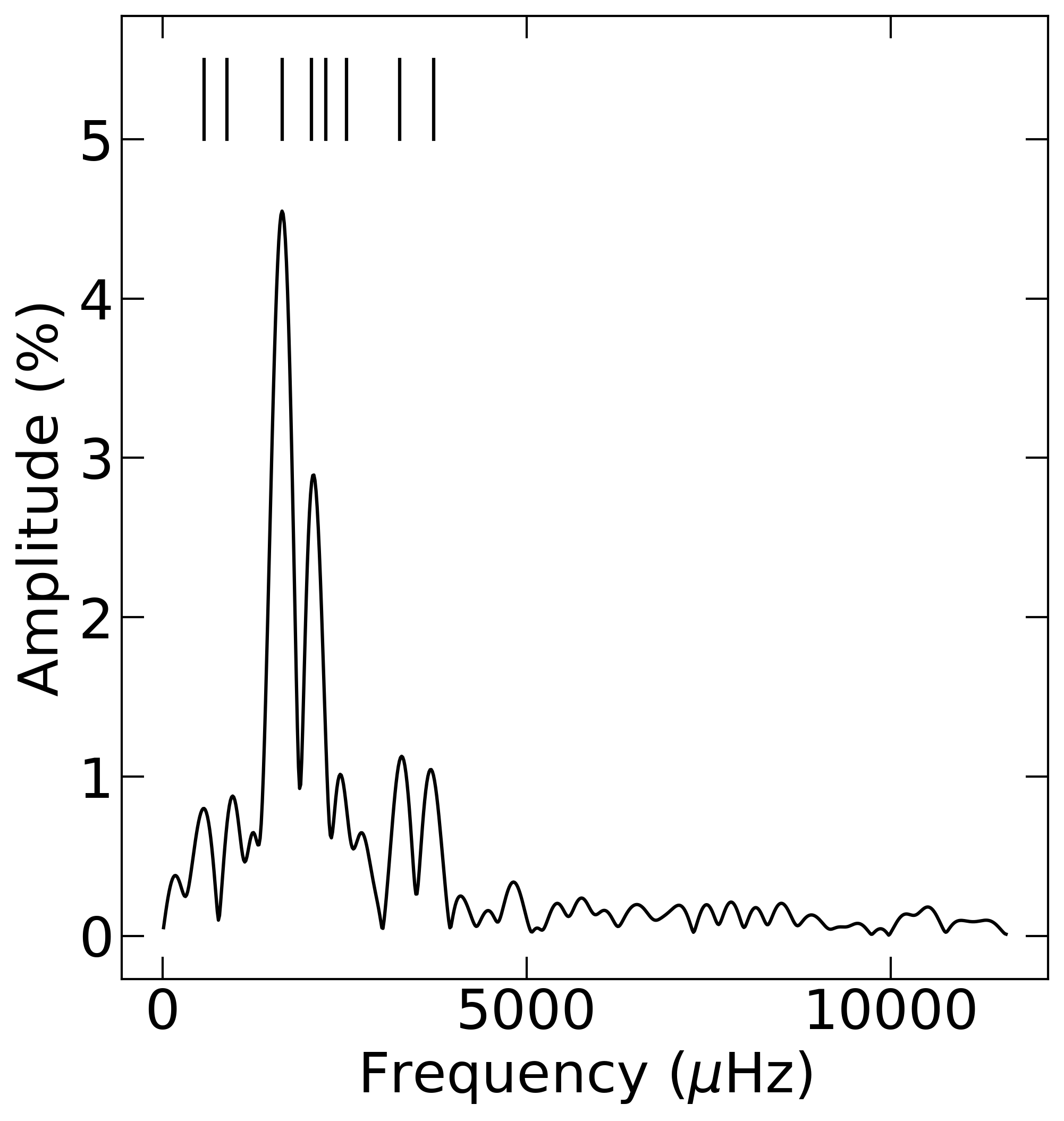}
\caption{The left panel shows an example of a light curve of a pulsating white dwarf. The data are in black, and a model fit is shown in red. The right panel shows the Fourier transform of the data, with the frequencies used in the model indicated by vertical bars.}
\label{fig:var}
\end{figure}

The first pulsating white dwarf to be discovered was HL Tau 76 in 1965, shortly followed by Ross 548 in 1970. Ross 548 was given the variable star designation ZZ Ceti, which is now used to refer to the class of hydrogen-atmosphere pulsating white dwarfs, also called DAVs. Since then, hundreds more were discovered and this is today the most populous class of pulsating white dwarfs. Pulsations in DAVs are observed to start around $12,000$~K, consistent with the partial ionization of hydrogen, and are observed to stop at $\approx 10,000$~K. Whereas it is understood why the pulsations start, the reason why they cease at $\approx 10,000$~K is not fully understood and an open subject of research. Such a range of temperatures where white dwarfs are observed to pulsate is known as an instability strip. The main white dwarf instability strips are illustrated in Fig.~\ref{fig:strips}. The amplitude of DAV pulsations is of typically no more than a few percent, and the periods are between 100~s and just under half an hour. An example of pulsator is shown in Fig.~\ref{fig:var}. At the very low mass end of the population ($\approx 0.2$~M$_{\odot}$), the pulsation periods are often longer (over an hour), reflecting the larger radii of these systems causing internal waves to take longer to propagate.

Unlike the serendipitous discovery of DAV white dwarfs, the variability of helium-atmosphere white dwarfs was predicted as a consequence of helium partial ionization in the early 1980s before being observed. Shortly thereafter, the first pulsating DB, GD~358, was discovered. It was given the variable designation V777~Her, which is another name for the DBV class. DBVs pulsate at higher temperatures than DAVs, typically between 22,000~K and 32,000~K. The amplitudes and periods of these pulsations are similar to those seen in DAVs. At even higher temperatures, there is another class of pulsating white dwarfs: DOVs or GW Vir stars (again, after the prototype of the class). They show pulsations for temperatures approximately in the range 80,000 to 180,000~K, with smaller amplitudes and longer periods (up to $\approx 2500$~s) than DAVs and DBVs.

Finally, there are other less numerous pulsation classes that are still under investigation: hot DAVs and DQVs. DQVs show variability between 19,000 and 22,000~K; however, the origin of their variability could be due to mechanisms other than pulsations. Hot DAVs have been predicted to exist with temperatures around 30,000~K. Very few such pulsators ($< 5$) have been detected to date, hence more observations are required to confirm the existence of this class.

Pulsating white dwarfs were originally discovered and characterized with targeted observations. Perhaps the most remarkable example are the observations made by the Whole Earth Telescope (WET), a network of telescopes around the world that allowed for continuous observations over long periods of time -- as the target set in one site due to sunrise, it could be observed from another site further West. These long-baseline observations allowed for studies of the stability of pulsation modes, revealing that amplitudes are often variable on timescales of days to years, but the observed characteristic pulsation modes remain the same. The pulsation modes do, however, show small changes on timescales of millions of years due to the residual contraction and cooling of the white dwarf, which slowly shifts its characteristic pulsations.

More recently, large photometric surveys, primarily designed for exoplanet research, have made significant contributions to the number of known pulsating white dwarf stars. These surveys can not only provide long baselines, like WET, but are also not subject to poor weather conditions like ground-based observations.

\section{The white dwarf cooling sequence and its substructures}

\begin{figure}[h]
\centering
\includegraphics[width=0.7\textwidth]{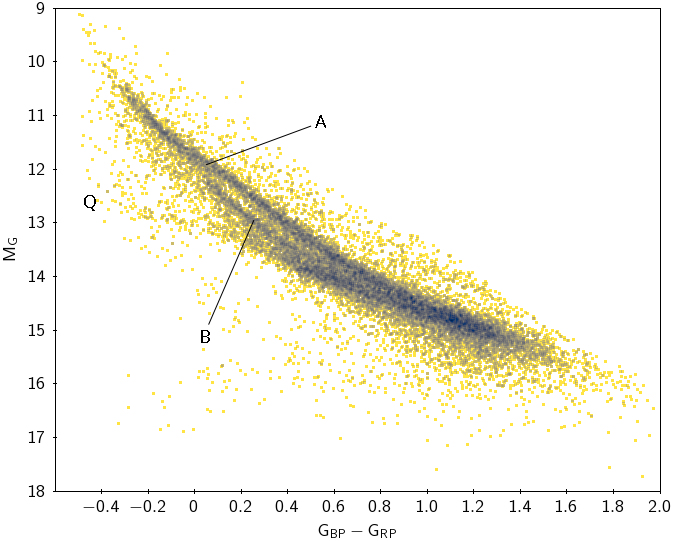}
\caption{The {\it Gaia} HR diagram showing in yellow the high-probability white dwarf candidates identified in {\it Gaia} data release 3. At least three different structures can be identified in the white dwarf sequence: the A, B, and Q branches.}
\label{fig:gaiahrwds}
\end{figure}

As well as allowing the identification of hundreds of thousands of white dwarf candidates, {\it Gaia} also provided a unique homogeneous and precise dataset that allowed an unprecedented view of the white dwarf cooling sequence \citep[as reviewed by][]{Tremblay2024}. In particular, it revealed the presence of different branches in the cooling sequence, that became known as A-, B-, and Q-branches (see Figure~\ref{fig:gaiahrwds}). The reason for this nomenclature is because, initially, it was believed that the A-branch was populated by DA stars, the B-branch by DB stars, and the Q-branch by DQ stars, though quickly it became clear that these assumptions were incomplete if not wrong.

The A-branch is indeed consistent with pure-hydrogen white dwarf stars. Pure-helium models for canonical mass white dwarfs, on the other hand, go through the gap in the cooling sequence, rather than tracing the B-branch. The proposed explanation has two aspects. First, the atmospheres are not entirely pure-helium, but rather still have traces of hydrogen, which is consistent with expectations from stellar evolution. The trace hydrogen displaces the helium cooling sequence downwards towards the B-branch; however, this alone would create a gradient rather than a gap. The gap can be explained by trace amounts of carbon being dredged-up from the core, which is known to happen \citep{Blouin2023}.

The Q-branch cannot be explained by an atmospheric composition effect. Instead, the accumulation of stars along the Q-branch is explained by another process that happens as white dwarfs cool down: crystallization. As the core temperature decreases, the ions in the core form a crystal and the associated release of latent heat leads to a cooling delay, implying that white dwarfs will spend a longer time at the crystallization temperature, causing an overdensity of stars in the range of temperatures corresponding to crystallization, which is precisely the one encompassed by the Q-branch. One caveat is that the cooling delay of crystallization alone is not long enough to explain the branch. One additional effect is the phase separation between carbon and oxygen in the core as the crystal forms, which is a consequence of the properties of a carbon-oxygen mixture: the solid will be oxygen-rich, whereas the liquid gets progressively more carbon rich. These two effects combined lead to a cooling delay of about 2~Gyr, but to fully explain the Q-branch a delay about five times longer is needed. The additional delay is a subject of discussion, with the distillation of $^{22}$Ne being the likely culprit. This heavy isotope is formed during stellar evolution and traces of it would be present in the core of the white dwarf. The carbon-oxygen crystals would be poor in neon when they form and would float upwards and eventually melt, whereas the neon enriched liquid is displaced downwards (distilled). When enough neon is present, this can lead to a delay of up to 10~Gyr. If a fraction of the white dwarf population ($<10$\%) is enriched in neon and experiences this additional delay, this would be enough to fully explain the Q-branch \citep{Bedard2024}.

A fourth, less prominent, branch can be seen in the range $16 < M_\mathrm{G} < 17$ and $-0.3 < G_\mathrm{BP} - G_\mathrm{RP} < 0.6$. This infrared faint sequence, sometimes referred to as ultrablue white dwarfs, arises as a result of collision induced absorption (CIA) from molecular hydrogen (H$_2$), in particular H$_2$-He collisions. Therefore only a fraction of white dwarfs, those with mixed hydrogen and helium atmospheres, turn infrared faint as they cool down.

\section{White dwarf stars in binary systems}

As previously mentioned, a large fraction of stars are in binary systems, with the fraction varying from about 20\% for low-mass stars to over 80\% for massive stars. Though some of these systems will merge during stellar evolution, many white dwarf stars are still observed as binary systems \citep[see review by][and other chapters in this Encyclopedia]{Charles2002}. These systems are of great importance in astronomy for three main reasons. First, binary interaction is still a poorly understood process -- modeling the stellar populations resulting from binaries, including white dwarf systems, can help constrain binary evolution models and improve our understanding of how stars exchange mass and angular momentum during evolution. Second, binary white dwarf stars are believed to be the progenitors of Type Ia supernovae (SN Ia). This class of supernovae qualifies as what is known as a standard candle: their intrinsic brightness can be derived from observed properties and, as a result, their distance can be inferred from the observed brightness. They are thus widely used for measuring cosmological distances and have been crucial for measuring the accelerated expansion of the Universe (2011 Nobel Prize in Physics). However, the exact mechanism behind SN Ia is not understood. Although it is widely accepted that they are associated with a thermonuclear explosion of a white dwarf that has approached the Chandrasekhar mass limit, observations have so far not found enough systems whose properties are consistent with the theoretical expectations of a SN Ia progenitor to explain the expected SN Ia rate of our Galaxy. This is known as the SN Ia progenitor problem. Finally, white dwarf binaries are the dominant source of low-frequency gravitational waves in our Galaxy. Although their signal is presently not directly detectable, the signal of future missions like the Laser Interferometer Space Antenna (LISA) is expected to be dominated by white dwarf systems. Therefore, to correctly process and interpret the signal to be detected by these missions, a good understanding of the population of white dwarf binaries is required.

Depending on the type of companion system, white dwarf binaries can be classified into two broad classes: single degenerate and double degenerate systems. In the former, only one component in the binary is a white dwarf, whereas double degenerate systems are composed of two degenerate objects. Both types of systems can be further classified into detached or interacting. In detached systems, there is no ongoing mass transfer between the two stars. In interacting systems, either one star (semi-detached system) or both stars (contact system) fill their Roche lobe -- the maximum radius around a star where its mass remains bound. This leads to mass being transferred from one star to the other.

\subsection{Detection of white dwarf stars in binaries}

From an observational point of view, binary systems can be either resolved or unresolved. Resolved systems, also called visual binaries, are those in which the two stars can be individually detected. One prominent example is Sirius, the brightest star in the night sky, which can be resolved into two components using magnification: a main sequence star (Sirius A) and a white dwarf (Sirius B, the second white dwarf to be discovered). Some resolved binaries are far enough apart that magnification is not required, and their association is instead inferred by their motion on the sky being consistent with being gravitationally bound to each other. These are also known as common proper motion pairs, since their consistent motion on the sky (proper motion) is what shows that they are bound. 

Unresolved binaries are the systems where the two stars are too close to each other to be individually resolved. They can be detected via astrometry, photometry, or spectroscopy. Astrometry can reveal the presence of a binary companion when the position of the visible star is observed to oscillate as the star orbits the center of mass of the system. Very precise positional measurements are required to detect this kind of system. Photometry can reveal binary systems in single measurements when the two stars contribute to the SED. A typical example are white dwarfs with M-dwarf companions: the white dwarf dominates the SED in the blue, whereas the M-dwarf dominates in the red. Similarly, this can also reveal binaries in spectroscopy when features of both stars are seen. If spectral lines of both stars are detected in spectroscopy, the system is called a double-lined binary.

\begin{figure}[h]
\centering
\includegraphics[width=0.8\textwidth]{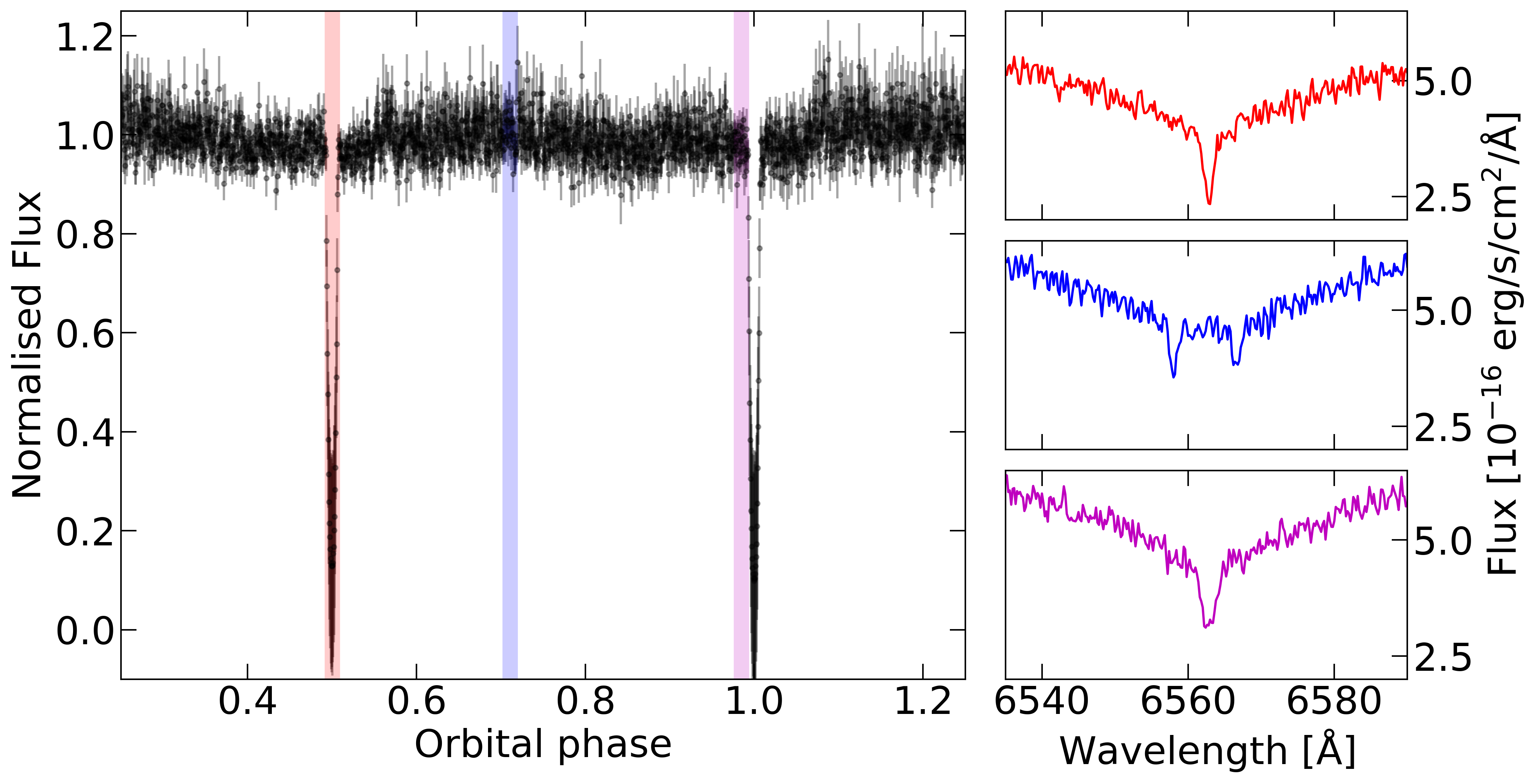}
\caption{The left panel shows the light curve of an eclipsing double white dwarf binary. Some orbital phases are marked by vertical bars. The spectra for these phases are shown on the right (left to right phases are shown from top to bottom). Around phase 0.75, where the separation between the two stars is maximized, the absorption lines of both stars can be seen. Near the eclipses, the two absorption lines combine into one visible feature.}
\label{fig:bin}
\end{figure}

The most efficient way of detecting binary systems is with time-series measurements, i.e. repeat measurements over time, which can reveal variations caused by a binary companion. Time-series photometry of binary white dwarf systems can reveal variability due to three main effects: eclipses, irradiation, and tidal deformation. In eclipsing binaries, flux is observed to drop when one star passes in front of the other. The orbital inclination of the system needs to be closely aligned with the line of sight, hence only a small fraction of binaries are observed to eclipse. Irradiation occurs when one star in the system is significantly hotter than the other, and can thus heat up one face of the cooler star. As a result, the temperature becomes inhomogeneous over the surface of the cooler star, and variable flux is detected as the system rotates. Finally, in systems with small orbital separation, tidal effects can lead to the deformation of one or both components away from spherical shape. This leads to a variable observable surface area as the system rotates, causing variability. Time-series photometry is only sensitive to relatively short orbital periods (typically less than a day) where these effects are noticeable. Time-series spectroscopy is in turn more sensitive and can be used to detected binaries in much wider orbits of hundreds or even thousands of days. Spectroscopy can reveal binaries due to the Doppler effect: the wavelength of observed spectral lines will change as stars orbit the center of mass, deviating to the blue when the star is moving towards the observer and to the red when its moving away. This is the same radial velocity method applied in the detection of extrasolar planets. An example white dwarf binary identified via time-series measurements is shown in Fig.~\ref{fig:bin}

Given the relative faintness of white dwarf stars compared to other stellar populations, the observed population of white dwarf binaries is strongly biased towards double degenerate systems and white dwarf stars with cool companions (typically of M-type), which have comparable luminosity to white dwarfs. Finding white dwarf companions to early-type main sequence, giant, or subgiant stars is a challenge. One of the methods applied is self-lensing: when the white dwarf passes in front of its bright companion, it may cause an increase in brightness as the white dwarf works as a lens that focuses the light in the direction of the observer. A handful of white dwarf companions to bright stars have been identified in this manner. A more successful method relies on finding excess flux in the ultraviolet in the SED of main sequence, giant, or subgiant stars, as white dwarfs can contribute to or even dominate the flux in the ultraviolet despite their small size, especially when they are young and hot. The radial velocity method is also applicable, but more time consuming as it requires repeated observations.

\subsection{Cataclysmic variables}

Systems where a white dwarf accretes mass from a cool companion are also known as cataclysmic variables \citep[reviewed by][]{Warner1995}. In these systems, mass accretion typically occurs via a disc, and variations in the brightness of this disc due to changes in accretion rate or disc dynamics lead to stochastic flux variability. These systems can also experience sudden and intense flux changes -- outbursts -- resulting from disc instabilities, which is what gives them the name cataclysmic.

Cataclysmic variables can be broadly classified based on the occurrence of outbursts, or lack thereof. Systems that exhibit outbursts are called dwarf-novae. Systems with high mass transfer rates but no observed outbursts are called nova-likes. Finally, some cataclysmic variables show brightness changes of over six magnitudes, which are attributed to a thermonuclear explosion on the surface of the white dwarf -- these are called classical novae.

In some cataclysmic variables, the presence of a magnetic field prevents or disrupts the formation of an accretion disc, as it establishes a preferential motion direction for the transferred matter. For magnetic fields up to 10~MG (1000~T), the external part of the disc can still form, but the internal disc is disrupted and accretion occurs along the magnetic field lines to the pole of the white dwarf. These systems are called intermediate polars. For fields stronger than that, the magnetosphere of the white dwarf fully encompasses the system and no disc forms -- mass transfer fully occurs along the magnetic field lines. These systems are known as polars.

The typical orbital periods of cataclysmic variables vary from just under 6~hours (beyond that, the system is detached) to a period minimum of around 80~min. The evolution of the orbital period is governed by two main processes: gravitational wave emission and magnetic braking, both of which cause the orbital period to slowly decrease as the binary evolves. Magnetic braking is the loss of angular momentum caused by material that is accelerated and ejected along the magnetic field lines. The efficiency of magnetic braking significantly drops for orbital periods around 3~h, when the companion star becomes fully convective and the mechanism behind its magnetic field changes. At this point, the companion will contract enough that it will not fill its Roche lobe and mass transfer stops. The system continues to contract due to gravitational wave emission and the cool companion fills its Roche lobe again when the period is around 2~h. This leads to a dearth of cataclysmic variables with orbital periods between 2 and 3 hours, the so-called period gap. 

The period minimum of 80~min is a direct consequence of the evolution of cataclysmic variables: as the cool companion loses mass, it will eventually fall below the hydrogen burning limit and become the equivalent of a brown dwarf. The response of a brown dwarf to mass loss is different from that of a main sequence star: it does not decrease in radius, and the radius can in fact even increase as mass decreases (as a consequence of different hydrostatic equilibrium condition, similar to white dwarfs). For Roche-lobe filling systems, the orbital period and the density of the mass-transferring star are related such that the decrease in density of the brown dwarf (as it both loses mass and its radius increases) leads to an orbital period increase. Systems that are in the stage of increasing orbital period are known as period bouncers. Observationally, there is a large discrepancy in the number of predicted and observed period bouncers which suggests some missing understanding of the evolution of cataclysmic variables.

\subsection{Pulsars and propellers}

There are two types of white dwarf binaries which present distinctive characteristics that set them aside from otherwise similar cataclysmic variable systems. These are the magnetic propellers and binary white dwarf pulsars.

In magnetic propellers, a companion is transferring mass to a magnetic white dwarf, but the white dwarf rotates at such a high rate that the material is accelerated and ejected from the system before being accreted. The two known magnetic propellers are AE~Aquarius, whose white dwarf has a rotation period of 33~s, and LAMOST J024048.51+195226.9, which is the confirmed white dwarf with the fastest rotation period: only 25~s.

Binary white dwarf pulsars, as the name suggests, present pulsing variability similar to the traditional neutron star pulsars. The mechanism is however not the same, and the pulses are believed to result from interaction between the magnetic fields of the white dwarf and of its companion. The pulsing behavior has been observed throughout the electromagnetic spectrum, from radio to X-rays. Like propellers, binary white dwarf pulsars show no evidence for accretion onto the white dwarf and are likely going through a brief detached phase as spin angular momentum is being transferred to the orbit and increasing the system separation. The two known systems are AR~Scorpii and J191213.72-441045.1, whose white dwarfs have rotation periods of 2 and 5~min, respectively.

\section{Evidence of planetary bodies around white dwarfs}

We now know of more than 5000 planets around main sequence stars. The vast majority of these stars will eventually become white dwarfs, which raises the question of the fate of these planetary systems during stellar evolution. When stars leave the main sequence and climb the giant branch, the increase in radius will cause the destruction of nearby planets, out to a few astronomical units. Planetary bodies with larger orbital separations, on the other hand, can survive. Gravitational interaction between the remaining bodies can push them to orbits closer to the white dwarf. At very close orbits (with periods around 4 hours), tidal forces are strong enough to destroy the planetesimal. This evolutionary process implies that a few different signatures of planetary systems can be found around white dwarfs: planetary debris, gas and dust discs, or surviving intact planets \citep[see][for additional details]{Veras2021}.

\begin{figure}[h]
\centering
\includegraphics[width=\textwidth]{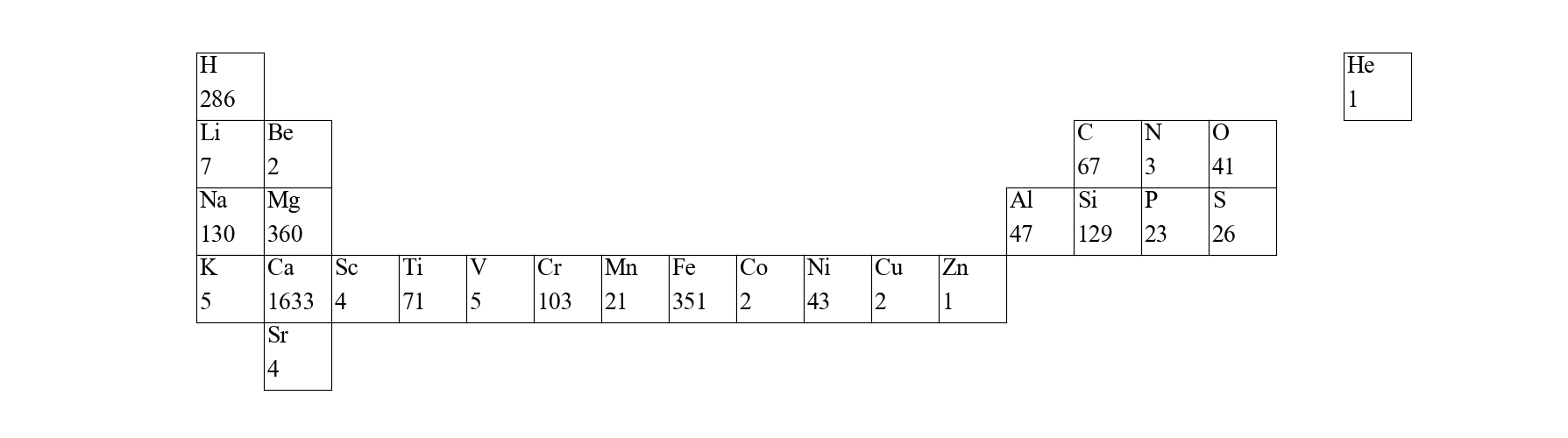}
\caption{The periodic table of chemical elements found in metal-enhanced white dwarfs. The number in each cell indicates in how many white dwarfs traces of that element linked to accretion were found as of April 2024.}
\label{fig:ptable}
\end{figure}

Planetary debris are the most common observed signature: the presence of metals in the atmosphere of white dwarf stars (the D[...]Z class) is an indirect evidence of debris orbiting and falling into the white dwarf. That is because the sinking time (or diffusion timescale) of metals is very short due to the high gravity of white dwarfs, such that any metals presently being observed must be accreting currently or very recently onto the white dwarf. Over 10\% of white dwarf stars show metal enhancement consistent with this scenario. Curiously, the third white dwarf to be discovered, van Maanen 2, is one example, although this was not realized until almost a hundred years after its discovery. In the optical, the most commonly detected metal is calcium, followed by magnesium, iron, and sodium. Overall, more than 1500 white dwarfs have been analyzed in detail to determine metal abundances, and a large number of elements have been detected, including lithium, beryllium, nitrogen, oxygen, aluminum, silicon, phosphorus, sulfur, and potassium. The complete list of detected elements and their occurrence is illustrated in Figure~\ref{fig:ptable}. The overall abundances that are detected are similar to those of rocky bodies, including planets, and known comets and asteroids, supporting the planetary hypothesis. More specifically, the material shows a variety of relative abundances that are consistent with meteorites, with bulk Earth, with core-rich, mantle-rich, or crust-rich material, with hydrated bodies, with icy volatile-rich objects, and with gas giant atmospheres.

\begin{figure}[h]
\centering
\includegraphics[width=0.7\textwidth]{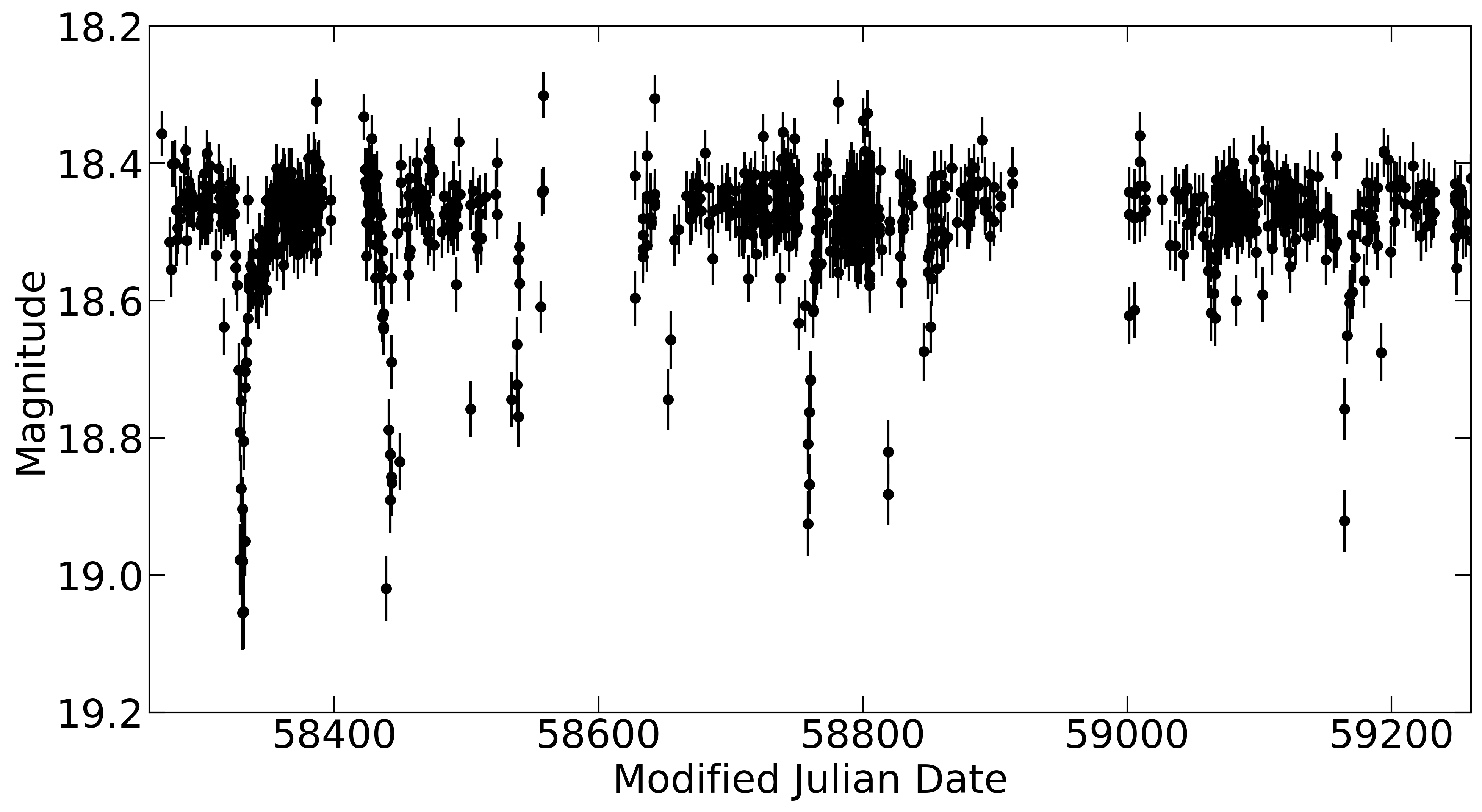}
\caption{Magnitude as a function of time for a white dwarf showing transiting debris, which is what causes the sharp drops in brightness.}
\label{fig:transit}
\end{figure}

Another method for the detection of debris is time-series photometry. Debris transiting in front of the white dwarf leads to a flux decrease called a transit, as illustrated in Figure~\ref{fig:transit}. The first white dwarf to exhibit such a feature was WD1145+017, which showed transits with periods between 4.5 and 4.9~h, consistent with a body destroyed by tidal forces. However, since then, seven other white dwarfs have showed evidence for transiting debris and they show a variety of periods -- sometimes much longer than expected if the orbiting debris were indeed a result of tidal disruption, showing that our understanding of how debris form and are scattered into the white dwarf is still incomplete.

As well as transiting debris, WD1145+017 also shows an excess of infrared flux compared to what is expected from a white dwarf, which is explained by a dusty debris disc around this system. A few dozen other white dwarfs have been found to show this same feature. In WD1145+017 and in a subset of systems showing dust discs, metallic emission lines indicate the presence of gaseous debris disc where gas is produced via ongoing collisions. In one particular case, WD J0914+1914, the observed composition of the gas indicates the presence of an evaporating giant planet in a close-in orbit.

Despite this abundance of indirect evidence for planetary bodies around white dwarfs, direct detection of planets is a challenge. The two most successful methods for planet detection around main sequence stars, transit and radial velocities, are not very efficient for white dwarfs. The transit method, whereby dips in flux caused by the planet passing in front of the star are used to infer its existence, is made difficult by the fact that white dwarfs are very small, which reduces the transit probability. One planet has been detected using this method, WD 1856+534b. The radial velocity method in turn requires measuring the motion of the star caused by the gravitational pull of the planet using the resulting Doppler shift of its spectral lines. However, the spectroscopic lines of white dwarfs are broader than of main sequence stars, as a result of the high gravity, leading to less precise (by at least a factor of three) radial velocities. No planet has been detected around white dwarfs using radial velocities as of yet. One planet, PSR B1620-26ABb (or PSR B1620-26c), has been detected via the pulsar timing method, where changes in the period of the regular pulses of a rapidly spinning neutron star (a pulsar) indicated the presence of another body whose gravity alters the frequency of pulses. In the case of PSR B1620-26, the timing has revealed that the pulsar (PSR B1620-26A) has a white dwarf companion (PSR B1620-26B), and that there is a planet orbiting both stars (a circumbinary planet), initially dubbed PSR B1620-26c, and more recently PSR B1620-26ABb to indicate its circumbinary nature. One planet candidate has been found through direct imaging, WD 0806-661b; however, the configuration of the system leaves open the possibility that this source is a brown dwarf rather than a planet. Finally, the planet MOA-2010-BLG-477Lb was discovered via the microlensing technique, where the light of a distant star is magnified when another object (the lens) passes in front of it from our perspective. The pattern of the magnification will be symmetric if the lens is a single star, but will be distorted if the lens is a binary, or if it has planets. For MOA-2010-BLG-477, the pattern of distortion suggests a low-mass star ($\approx 0.5$~M$_{\odot}$) and a planet with a mass about 50\% higher than Jupiter. The low-mass star was not detected directly, implying it is a dim white dwarf.

\section{Volume-limited white dwarf samples}

A major disadvantage of earlier white dwarf samples, in particular those compiled from spectroscopic surveys like SDSS, is that they were limited by magnitude. This leads to selection biases implying that the observed population does not reflect the underlying population. For example, with the blue-targeted selection of SDSS, cool and red white dwarfs -- which are generally higher mass systems, as they are older and therefore formed first -- were underrepresented in the observed population.

The way to avoid these biases and to obtain an accurate observational picture of the white dwarf population is to make use of volume-limited samples: rather than selecting on magnitude, selecting on distance, defining a volume around the Earth where all white dwarf stars are accounted for. The first attempt at obtaining a volume-limited sample started with a 13~pc sample, which contained only 51 systems \citep{Holberg2002}. The problem then becomes another: low-number statistics, resulting in rarer systems being underrepresented or completely absent from the sample. This sample was first extended to 20~pc \citep[139 systems, but then believed to be only 80\% complete;][]{Holberg2008} and then to 25~pc \citep[232 systems, then 68\% complete;][]{Holberg2016}.

{\it Gaia} enabled an increase in the completion volume significantly due to its improved sensitivity, allowing the compilation of a complete 40~pc volume-limited sample with over a thousand white dwarf systems \citep{OBrien2024}. With this sample, it is possible to quantify the percentage of white dwarf systems within the different spectroscopic classes, as well as the fraction of binaries.

Concerning spectroscopic classes, it is found that more than 60\% of white dwarf stars are of type DA. The next most abundant class is DCs, with over 25\%. DOs and DBs are not abundant in this sample (0 and 2\%) because most stars in this local sample are expected to be old enough to have cooled down below the helium excitation threshold and become DCs. Around 4\% of the sample are DQs. Metal enhancement is observed for 11\% of the sample; however, the detection of metals is limited by the quality of data, hence this is likely a lower limit. Similarly, magnetic fields are detected for 7\% of the sample, but more detailed studies suggest that a fifth of white dwarf stars are magnetic. The fraction of binaries is around 20\%, similar to what was found for the smaller volume samples, but slightly lower than model predictions. Whether this is due to limitations in the observational detection of binaries, or an issue with the models, remains to be determined.

\section{Summary and Outlook}

White dwarf research has made extraordinary progress in recent decades thanks to large spectroscopic, photometric, and -- importantly -- astrometric surveys. These surveys have revealed a variety of spectral types and hinted at the occurrence of internal phenomena that enabled a better understanding of white dwarf evolution and of their interaction with the surrounding environment, in particular planetary bodies. In the future, new spectroscopic surveys such as SDSS V, the Dark Energy Spectroscopic Instrument (DESI), the 4-metre Multi-Object Spectroscopic Telescope (4MOST), and the William Herschel Telescope Enhanced Area Velocity Explorer (WEAVE) will increase the number of spectroscopically confirmed white dwarfs by an order of magnitude, potentially revealing new features and allowing additional insight into small but growing populations. Photometric surveys like those carried out by the Vera C. Rubin Observatory and the Nancy Grace Roman Space Telescope will provide photometry of fainter white dwarf stars in underexplored regions of the Galaxy, enabling new research into white dwarfs with different progenitor populations. Perhaps most exciting of all, LISA might directly detect the gravitational waves from white dwarf binary systems for the first time, opening a new window into white dwarf research and bringing white dwarf stars to the forefront of multi-messenger astronomy.

\begin{ack}[Acknowledgments]

Ingrid Pelisoli acknowledges support from a Royal Society University Research Fellowship (URF\textbackslash R1\textbackslash 231496). We thank Antoine B\'{e}dard, Andy Buchan, Boris Gaensicke, James Munday, Dimitri Veras, and Pier-Emmanuel Tremblay for providing comments on an earlier version of this manuscript.

\end{ack}

\seealso{article title article title}

\bibliographystyle{Harvard}
\bibliography{reference}

\end{document}